\documentclass[aps,prd,reprint,amsmath,amsfonts,nofootinbib,floatfix,superscriptaddress]{revtex4-2}

\usepackage[utf8]{inputenc}
\usepackage[dvipsnames]{xcolor}
\usepackage{graphicx}
\graphicspath{{plots/}}
\usepackage[colorlinks=true,     % use colored text instead of boxes
            linktoc=all,     %set to all if you want both sections and subsections linked
    	linkcolor=NavyBlue,  %choose some color if you want links to stand out
    	citecolor=NavyBlue,
         urlcolor=cyan    % color of URLs (including in bibliography)
           ]{hyperref}
\usepackage{url}
\usepackage{siunitx}

\hypersetup{
    colorlinks=true, %set true if you want colored links
    linktoc=all,     %set to all if you want both sections and subsections linked
    linkcolor=NavyBlue,  %choose some color if you want links to stand out
    citecolor=NavyBlue,
}

\newcommand{\sub}[1]{_{\mathrm{#1}}}
\newcommand{\super}[1]{^{\mathrm{#1}}}
\newcommand{\Romer}{\text{R\o{}mer}}
\newcommand{\prob}[2]{P\left(#1|#2\right)}
\newcommand{\scalar}[2]{\left(#1|#2\right)}
\newcommand{\Gauss}{\mathcal{P}}
\newcommand{\avg}[1]{\left\langle #1 \right\rangle}
\newcommand{\dfact}{\gamma}
\newcommand{\As}{A\sub{s}}
\newcommand{\Ac}{A\sub{c}}
\newcommand{\xs}{x\sub{s}}
\newcommand{\xc}{x\sub{c}}
\newcommand{\F}{\mathcal{F}}
\newcommand{\Fphi}{\F_\phi}
\newcommand{\expect}[1]{E\left[#1\right]}
\newcommand{\Ord}[1]{\mathcal{O}\left(#1\right)}
\newcommand{\dop}{\lambda}

\newcommand{\software}[1]{\textsc{#1}}
\newcommand{\code}[1]{\texttt{#1}}
\newcommand{\vn}{\hat{n}}
\newcommand{\vecr}{\vec{r}}
\newcommand{\vx}{\vec{x}}

\newcommand{\vro}{\vec{r}\sub{o}}
\newcommand{\ro}{r\sub{o}}
\newcommand{\zo}{z\sub{o}}
\newcommand{\zs}{z\sub{s}}
\newcommand{\vB}{\mathbf{B}}

\newcommand{\src}{\sub{src}}
\newcommand{\truesrc}[1]{\check{#1}}
\newcommand{\lalsuite}{\software{LALSuite}}
\newcommand{\pint}{\software{Pint}}
\newcommand{\tempo}{\software{tempo}}
\newcommand{\tempoTwo}{\software{tempo2}}
\newcommand{\astropy}{\software{Astropy}}
\newcommand{\AU}{\mathrm{AU}}
\newcommand{\abs}[1]{\left|#1\right|}
\newcommand{\sun}{_{\odot}}
\newcommand{\dtau}{\delta\tau}
\newcommand{\stddtau}{\sigma_{\dtau}}
\newcommand{\ns}{\nano\second}
\newcommand{\geo}{\mathrm{geo}}
\newcommand{\obs}{\mathrm{obs}}
\newcommand{\vrobs}{\vec{\mathfrak{r}}\sub{obs}}
\newcommand{\asini}{a\sub{p}}
\newcommand{\porb}{P\sub{B}}
\newcommand{\Rorb}{R\sub{B}}
\newcommand{\ecc}{e}
\newcommand{\argp}{\omega}
\newcommand{\tperi}{t\sub{p}}
\newcommand{\oldTDB}{TDB$\sub{old}\super{TAI}$}
\newcommand{\newTDB}{TDB$\sub{new}\super{TAI}$}
\newcommand{\newTDBbipm}{TDB$\sub{new}\super{BIPM}$}
\newcommand{\TT}{\text{TT}}
\newcommand{\TAI}{\text{TAI}}
\newcommand{\GPS}{\text{GPS}}
\newcommand{\BIPM}{\text{BIPM}}
\newcommand{\DD}{$\text{DD}\sub{BT}$}
\newcommand{\cosi}{\cos\iota}

\begin{document}
\title{Validating Timing-Model Accuracy for Continuous Gravitational Waves: A Comparison of LALSuite and PINT}

\author{Kartikey Sharma}
\author{Reinhard Prix}
\author{Maria Alessandra Papa}
\affiliation{Max Planck Institute for Gravitational Physics (Albert Einstein Institute), D-30167 Hannover, Germany}
\affiliation{Leibniz University Hannover, D-30167 Hannover, Germany}
\author{Curt Cutler}
\affiliation{Jet Propulsion Lab, California Institute of Technology, 4800 Oak Grove Drive, Pasadena, California 91109, USA}
\affiliation{Theoretical Astrophysics, California Institute of Technology, Pasadena, California 91125, USA}

%\date{\commitDATE; \commitIDshort-\commitSTATUS}

\begin{abstract}
  We present results of a systematic validation of the \lalsuite{} timing model for continuous gravitational waves
  against \pint{}, a modern high-accuracy pulsar-timing package.
  An accurate timing model is essential for tracking the signal phase, and hence for detecting and accurately
  characterizing continuous gravitational waves.
  In order to quantify the impact of timing inaccuracies, we derive and validate the leading-order relation
  $\mu \approx (2\pi f)^2\stddtau^2$, where $\mu$ is the fractional loss of signal power, $f$ is the signal frequency,
  and $\stddtau^2$ is the variance of the timing errors.
  We then compare the solar-system and binary components of the \lalsuite{} timing model against the corresponding models
  in \pint{}.
  With the original \lalsuite{} Einstein-delay implementation, the total disagreement is dominated by that component and
  has $\stddtau\simeq\SI{2.3}{\micro\second}$ (corresponding to $\mu\simeq\SI{0.02}{\percent}$ at $f=\SI{1000}{\hertz}$).
  With the newer Einstein-delay implementation, the total disagreement (over one year) drops to
  $\stddtau\lesssim\SI{31}{\nano\second}$ (or $\mu\lesssim\num{4e-8}$ at $f=\SI{1000}{\hertz}$) and is dominated by the
  observatory contribution to the \Romer{} delay, owing to the approximate Earth-rotation model used by \lalsuite{}.
  We additionally test binary delays using orbital parameters from \num{474} catalogued binary pulsars and verify the
  self-consistency of the \lalsuite{} source-time derivatives.
  Finally, we derive and discuss the \lalsuite{} Shapiro delay for signals passing through the solar interior, a case
  only relevant to gravitational waves.

\end{abstract}

\maketitle

\section{Introduction}
Continuous gravitational waves (or \emph{continuous waves} for short) are long-lasting, quasi-monochromatic signals.
The amplitudes of such continuous waves are several orders of magnitude lower than those of the typical transient
signals from the merger of compact binaries that current ground-based detectors routinely observe
\cite{LIGOScientific:2025slb}.

Searches for continuous waves have been an area of active research for the past two decades, e.g., see recent examples
of targeted searches \cite{abac_search_2025,mcgloughlin_search_2026}, directed searches
\cite{ming_observational_2026,collaboration_searches_2026} and all-sky searches
\cite{covas_wide_2026,collaboration_all-sky_2026}, as well as the reviews \cite{Riles:2022wwz,wette_searches_2023}.
The first detection of continuous waves will be a milestone event, unlocking gravitational-wave astronomy beyond the
transient sky and opening a new window into the extreme-matter regime inside neutron stars \citep{Lu:2023byl}.

The primary sources for ground-based detectors are expected to be rapidly rotating neutron stars with deviations from
perfect axisymmetry (e.g., see \cite{Gittins:2021zpv,Morales:2023euv} and references therein).
Despite steady improvements in search breadth and depth, with current searches increasingly probing plausible
neutron-star deformations, no continuous-wave signal has yet been detected.

While the absence of a detection is consistent with predicted source populations lying below current sensitivity levels
(e.g., \citep{Pagliaro:2023bvi}), it is important to examine and validate all elements of our search methods and
implementations.
In this paper we focus on one particularly critical element: \emph{the timing model}.

The timing model relates detector-frame arrival times to effective source-frame times by accounting for the relevant
time-dependent delays.
Long integrations require accurate tracking of the signal phase, while a time-dependent timing error produces a phase
error (proportional to signal frequency).
Timing accuracy is therefore essential for detecting and accurately characterizing such signals.

Every continuous-wave search method relies on a timing model, and many of the current search pipelines use the timing
code in \lalsuite{} \cite{lalsuite}.
Precision pulsar timing relies on the same barycentric and binary time-delay models, with accuracy requirements at least
as stringent as those of continuous-wave searches.
This is why we have chosen \pint{} \cite{pint}, a modern Python-based pulsar-timing package, as an independent
high-accuracy reference implementation for our comparison.

The most recent published validations of the \lalsuite{} timing-model implementation (against \tempo~\cite{tempo}) date
back approximately two decades, namely \cite{LIGO-tempo} (Fig.~2), and \cite{pitkin_binary_2007} (Fig.~6)
for binary sources.
While there have been internal spot-checks since then, in light of the central importance of this part of the search
codes, it is timely and necessary to perform and document a systematic validation study.

The plan of this paper is as follows: in Sec.~\ref{sec:timing-model} we introduce the full continuous-wave timing model,
and in Sec.~\ref{sec:mismatchcalculation} we derive and validate an expression linking timing errors to loss of signal
power.
In Sec.~\ref{sec:comp-lals-with} we perform numerical timing-model comparisons against \pint{}, and
Sec.~\ref{sec:conclusions} gives our conclusions.
In Appendix~\ref{sec:solar-shapiro-delay} we derive the Shapiro delay for signals passing through the Sun.

\section{Timing model}
\label{sec:timing-model}

In this section we describe the timing model that the continuous-wave signal model is built upon, as well as its
implementation in \lalsuite{} \cite{lalsuite}.

\subsection{The signal model}
\label{sec:signal-model}

A continuous-wave signal in the detector can be written \cite{jks1} as
\begin{equation}
\label{eq:fullSignalModel}
	h(t) = F_{+}(\vn, \psi; t)\, h_{+}(t) + F_{\times}(\vn, \psi; t)\, h_{\times}(t),
\end{equation}
where $\vn$ is a unit-vector pointing to the source, $\psi$ is the gravitational-wave polarization
angle and $F_{+}$ and $F_{\times}$ are detector beam-pattern functions for the two polarizations, $+$ and $\times$.
The two polarization waveforms take the form
\begin{equation}
  h_{+}(t) = A_{+}\cos\phi(t), \quad
  h_{\times}(t) = A_{\times}\sin\phi(t),
\end{equation}
where $\phi(t)$ is the signal phase in the detector frame, and $A_{+}$ and $A_{\times}$ are the corresponding
polarization amplitudes, which can be further expressed in terms of an intrinsic gravitational-wave amplitude $h_0$ and
the source-inclination angle $\iota$ (with respect to $\vn$), namely
\begin{equation}
  A_{+} = \frac{1}{2}h_{0}(1 + \cos^{2}{\iota}), \quad
  A_{\times} = h_{0}\cos{\iota}.
\end{equation}
Continuous waves from spinning neutron stars are quasi-monochromatic, so the (true) source-frame phase can be written as
\begin{equation}
  \label{eq:20}
  \truesrc{\phi}(\truesrc{\tau}) = \truesrc{\phi}_0 + 2\pi \int_{\truesrc{\tau}_0}^{\truesrc{\tau}} \truesrc{f}(\tau')\,d\tau',
\end{equation}
in terms of a slowly-varying intrinsic frequency $\truesrc{f}(\tau)$.

\subsection{Time delays}
\label{sec:time-delays}

If we consider a wavefront phase $\phi(t)$ at the detector that was emitted by the source at time $\truesrc{\tau}(t)$,
then
\begin{equation}
  \label{eq:truephase}
  \phi\left(t\right)= \truesrc{\phi}\left(\truesrc{\tau}(t)\right).
\end{equation}
We define the \emph{time delay} $\truesrc{\Delta}$ between emission- and arrival time as
$\truesrc{\tau}(t) = t - \truesrc{\Delta}(t)$, using the sign convention\footnote{Following pulsar astronomy, e.g.,
  \pint{} Eq.~(2) in \cite{pint}, \tempoTwo{} Eq.~(8) in \cite{edwards_tempo2_2006}, while other sign conventions also
  exist in the literature, e.g., Eq.~(4) in \cite{KAGRA:2022dwb} or Eq.~(1) in \cite{pitkin_reduced_2018}.}
``delay $\equiv$ arrival-time - emission-time''.

This true delay can be separated into $\truesrc{\Delta} \equiv \Delta\sub{d} + \Delta$, where $\Delta\sub{d} \equiv d/c$
is the light-travel time over the distance $d$ between the solar-system barycenter (SSB) and the source-system
barycenter (which would be the source frame for isolated sources, or the binary-system barycenter for sources in
binaries).
The distance $d$ is generally time dependent and unknown, but we can absorb it by defining an \emph{effective} source
frame with phase $\phi\src(\tau)$, defined as
\begin{equation}
    \label{eq:21}
    \phi\src(\tau) \equiv \truesrc{\phi}\left(\tau - \Delta\sub{d}(\tau)\right)\,,
\end{equation}
where $\tau$ is the arrival time in the effective source frame of a wavefront emitted at time
$\truesrc{\tau}(\tau) = \tau - \Delta\sub{d}(\tau)$ in the true source frame.

We can therefore describe the wavefront $\phi(t)$ arriving at the detector as being emitted from the effective source
frame at time $\tau(t)$, namely
\begin{equation}
  \label{eq:4}
  \begin{aligned}
    \phi(t) &= \phi\src(\tau(t)),\quad\text{where}\\
    \tau(t) &= t - \Delta(t),
  \end{aligned}
\end{equation}
with an \emph{effective} time delay $\Delta(t)$, which is fully determined for a given sky position and binary-orbital
parameters, and can be separated into several contributions:
\begin{align}
  \Delta &\equiv \Delta\sub{\Romer} + \Delta\sub{Shapiro} + \Delta\sub{Einstein} + \Delta\sub{Binary}, \label{eq:19b}
\end{align}
namely the solar-system \Romer{} delay $\Delta\sub{\Romer}$, relativistic Shapiro delay $\Delta\sub{Shapiro}$, and
Einstein delay $\Delta\sub{Einstein}$, as well as a binary-orbital time delay $\Delta\sub{Binary}$ for sources in binary
systems.
Note that gravitational waves always travel at the speed of light, and are therefore not affected by interstellar medium
or atmospheric effects (contrary to electromagnetic waves), which is why the corresponding delay terms of the pulsar
timing model (e.g., see Eq.~(3) in \cite{pint}) are absent here.

The (effective) frequency- and spindown parameters ${f, \dot{f}, \ddot{f}, \ldots}$ of a continuous-wave signal are
defined in terms of a Taylor expansion of the (effective) source-frame phase $\phi\src$ around a reference time
$\tau_0$, namely
\begin{equation}
  \label{eq:22}
  \phi\src(\tau) = \phi_0 + 2\pi\left[f(\tau-\tau_0) + \frac{1}{2} \dot{f}(\tau-\tau_0)^2 + \ldots\right],
\end{equation}
which can differ from the true frequency- and spindown parameters due to relative velocity $\dot{d}$ and acceleration
$\ddot{d}$ between source-system and SSB frames, as observed for systems in globular clusters
\cite{ransom_pulsars_2007}, for example.

For performance reasons, the $\F$-statistic implementations in \lalsuite{} employ a piecewise-linear approximation to
the detector-frame phase $\phi(t)$ of Eq.~\eqref{eq:4}, by computing the time delay $\Delta(t)$ \emph{and its
  derivative} $\dot{\Delta}(t)$ in steps $t_{i+1}=t_i+T\sub{sft}$ with $T\sub{sft}\sim\Ord{60-1800\si{\second}}$,
and using linear interpolation in between, i.e., $\Delta(t) \approx \Delta(t_i) + (t - t_i)\dot{\Delta}(t_i)$ for
$|t-t_i|<T\sub{sft}/2$.
The step size $T\sub{sft}$ is chosen such that the maximal expected loss of signal power (see
Sec.~\ref{sec:mismatchcalculation}) due to this approximation is bounded below a certain value (by default
$\SI{1}{\percent}$), e.g., see Eq.~(C2) in \cite{LeaciPrixBinary}.

\subsubsection{Solar-system \Romer{} delay}
\label{sec:solar-system-romer}

The solar-system \Romer{} delay $\Delta\sub{\Romer}$ refers to the purely geometric (flat space-time) delay between a
wavefront reaching the SSB (at $t\sub{SSB}$) and arriving at the detector at time $t$, therefore
\begin{equation}
  \Delta\sub{\Romer} \equiv t - t\sub{SSB}(t) = - \frac{\vecr(t) \cdot \vn}{c},
  \label{eq:roemer}
\end{equation}
where $\vecr(t)$ is the instantaneous position of the detector with respect to the SSB, and we are neglecting
finite-distance parallax effects.

We can write the detector position as $\vecr = \vecr\sub{\geo}+\vrobs$, in terms of the Earth geocenter
$\vecr_{\geo}$ and the relative detector-position offset $\vrobs$, separating $\Delta\sub{\Romer}$ into
corresponding geocentric- and observatory contributions, i.e., $\Delta\sub{\Romer} = \Delta\sub{\Romer,\geo} +
\Delta\sub{\Romer,\obs}$.

The geocentric position $\vecr\sub{\geo}(t)$ in the SSB is computed from the JPL solar-system ephemerides \cite{jpl},
which are provided as Chebyshev polynomial fits.
These are first converted\footnote{Using the \lalsuite{} tool \code{lalapps\_create\_solar\_system\_ephemeris}, or the
  newer python package \code{solar-system-ephemerides}~\cite{pitkin_ephemerides}.}
to ephemeris files holding time-sampled position, velocity and acceleration tables for the Earth, sampled every
\qty{2}{hours} (and for the Sun, sampled every \qty{20}{hours}).
From these the instantaneous position of the Earth $\vecr\sub{\geo}(t)$ (and Sun $\vecr\sun(t)$) can be computed by
extrapolation from the closest sample point.

To compute the observatory offset $\vrobs(t)$ for a detector, the rotational state of the Earth at time $t$ is
approximated using an analytical Earth-rotation model \cite{1992esaa.book}, which also takes leap seconds into account.
Note, however, that the pulsar-astronomy codes \tempoTwo{} and \pint{} use the more accurate empirical IERS Earth
orientation data \cite{iers} instead.

\subsubsection{Solar-system Shapiro delay}\label{sec:shapiro}

The Shapiro delay $\Delta\sub{Shapiro}$ accounts for the relativistic delay incurred by the propagation of a signal
through the curved spacetime in the solar system.
For signal rays not traversing the Sun, the leading-order term for the Shapiro delay is
\begin{equation}
  \Delta\sub{Shapiro}\super{ext} = - \frac{2G M\sun}{c^3} \ln\left[\frac{\vro\cdot\vn + \abs{\vro}}{1\AU}\right],
  \label{eq:shapiro}
\end{equation}
where $M\sun$ is the solar mass, $\vro \equiv \vecr - \vecr\sun$ is the vector from the Sun $\vecr\sun$ to the observer
$\vecr$, and the argument in the log is normalized by a length scale of $1\AU$ (the average Earth-Sun distance),
consistently with \tempoTwo/\pint{} conventions.
The numerical scale of this delay is $2G M\sun/c^3\sim\SI{10}{\micro\second}$.

\lalsuite{} uses an approximate version of this expression, with $\vro$ pointing to the geocenter instead of the
observatory, i.e., $\vro \approx \vecr\sub{geo}-\vecr\sun$, which neglects the observatory offset $\vrobs=\vecr - \vecr\sub{geo}$.
Furthermore \tempoTwo{}/\pint{} include additional delay contributions from other solar-system bodies and higher-order
corrections for close-by sources, see Eq.~(32) in \cite{edwards_tempo2_2006}.

The solar-exterior Shapiro-delay expression of Eq.~\eqref{eq:shapiro} does not hold for rays traversing the Sun and
diverges for sources exactly centered behind the Sun, i.e.\ $\vro\cdot\vn \rightarrow -\abs{\vro}$.
This is not a concern for electromagnetic waves and is therefore ignored in \tempoTwo{}/\pint{}, but it does need to be
dealt with for gravitational waves.
We define the relative impact parameter $b$ as,
\begin{equation}
  \label{eq:19}
  b  \equiv R\sun^{-1}\,\sqrt{\abs{\vro}^2 - (\vro\cdot\vn)^2},
\end{equation}
where $R\sun$ is the radius of the Sun.
For a ray going through the Sun, namely $b<1$ and $\vro\cdot\vn < 0$, \lalsuite{} implements the regular solar-interior
expression
\begin{equation}
  \label{eq:shapiro-interior}
  \begin{aligned}
    \Delta\sub{Shapiro}\super{int} &= \frac{4G M\sun}{c^3}\left(1 - b\right) + \Delta\sub{Shapiro}\super{ext,\mathit{b=1}}\,,
  \end{aligned}
\end{equation}
where the second term is the external Shapiro delay of Eq.~\eqref{eq:shapiro} evaluated for a grazing ray at $b=1$, by
substituting $\abs{\vro} = \sqrt{R\sun^2 + (\vro\cdot\vn)^2}$ in Eq.~\eqref{eq:shapiro}.
This expression was derived \cite{cutler_behind_sun} for a simple density model of the Sun.
In Appendix~\ref{sec:solar-shapiro-delay} we give its full derivation and compare it to a numerically-integrated
standard solar density model, which shows a maximal discrepancy of underestimating the Shapiro delay by about
$\sim\SI{14.5}{\micro\second}$ near the solar center.

\subsubsection{Solar-system Einstein delay}
\label{sec:einstein}

The Einstein delay $\Delta\sub{Einstein}$ refers to the time dilation experienced by terrestrial clocks due to the
Earth's orbital motion and the local gravitational potential (dominated by the Sun) with respect to an SSB time standard
such as Barycentric Coordinate Time (TCB) or Barycentric Dynamical Time (TDB) \cite{kaplan_iau_2006,klionertdb}.
Here we focus exclusively on TDB as the barycentric time standard, which is the default (and better supported) by both
\lalsuite{} and \pint{}, so the corresponding Einstein delay is defined as
\begin{equation}
  \Delta\sub{Einstein} \equiv t\sub{TT} - t\sub{TDB},
\end{equation}
where $t\sub{TT}$ is the time coordinate of an event (such as a wavefront arriving at the detector) measured in the
terrestrial-time (TT) standard, while $t\sub{TDB}$ is the barycentric (TDB) time coordinate of the same event.
The Einstein delay therefore only depends on the relative position and velocity of the detector in the solar system,
and is independent of the signal parameters.

We can separate $\Delta\sub{Einstein}$ into two contributions, similarly to \Romer{}-delay, namely
$\Delta\sub{Einstein}=\Delta\sub{Einstein,\geo}+\Delta\sub{Einstein,\obs}$, where $\Delta\sub{Einstein,\geo}$ is the
Einstein delay experienced at the center of the Earth $\vecr\sub{\geo}$, and
$\Delta\sub{Einstein,\obs}=-\vrobs\cdot\dot\vecr\sub{\geo}/c^2\sim\Ord{\SI{2}{\micro\second}}$ is the correction
due to the offset $\vrobs$ of the detector from the geocenter (cf.\
\cite{moyer_transformation_1981,edwards_tempo2_2006} and implementations in \tempoTwo{}, \pint{} and
\astropy{}~\cite{collaboration_astropy_2022}).

\lalsuite{} has two different implementations of $\Delta\sub{Einstein}$:
\begin{itemize}
\item the original \code{XLALBarycenterEarth} (C.~Cutler (2001)), which computes $\Delta\sub{Einstein,\geo}$ using the
  first $\num{20}$ terms (of about $\num{800}$ used by \tempo~\cite{tempo}) from an analytical series-expansion
  approximation \cite{fairhead}, and which neglects the observatory correction $\Delta\sub{Einstein,\obs}$.

\item the newer \code{XLALBarycenterEarthNew} (M.~Pitkin (2012)), which interpolates $\Delta\sub{Einstein,\geo}$ from
  numerically-computed \cite{irwinfukushima} \tempoTwo{} time-ephemeris files, and which includes\footnote{technically
    this step happens in \code{XLALBarycenter}()}
  the observatory contribution $\Delta\sub{Einstein,\obs}$.
\end{itemize}
We note that \lalsuite{} always uses the idealized $\TT(\TAI) = \TAI + \SI{32.184}{\second}$ realization of the TT time
standard (where $\TAI = \GPS + \SI{19}{\second}$), while \pint{} can optionally use the more accurate $\TT(\BIPM)$
realization\footnote{\url{https://www.bipm.org/en/time-ftp/tt-bipm-}}, which is empirically corrected for long-term
drifts in the atomic-time standard $\TAI$.

\subsubsection{Binary Delay}
\label{sec:binary-delay}

A signal emitted by a source in a binary system is subject to an additional time delay $\Delta\sub{Binary}$ due to
orbital motion and relativistic effects.
Here we focus on the $\F$-statistic-related code paths\footnote{The Glasgow known-pulsar pipeline
  \cite{pitkin_cwinpy_2022} implements more binary models, some of which are discussed in \cite{pitkin_binary_2007}.}
in \lalsuite{}, namely the function \code{XLALAddBinaryTimes()}, which only implements the Keplerian
Blandford~\&~Teukolsky (BT) model \cite{btmodel} with parameters $\dop\sub{B}$:
the orbital period $\porb$, the projected semi-major axis $\asini$ (measured in light-seconds), the eccentricity $\ecc$,
the time of periapsis passage $\tperi$, and the argument of periapsis $\argp$.
In this non-relativistic model the binary delay consists only of the orbital \Romer-delay, namely
\begin{equation}
  \Delta\sub{Binary} = \frac{\Rorb(\tau)}{c},
  \label{eq:binarydelay}
\end{equation}
where $\Rorb$ is the radial (i.e., along the line of sight) distance of the source with respect to the binary barycenter
(BB), at the emission time $\tau$.
Note that $\Rorb>0$ means the source is farther away from the observer than the BB.

Following \cite{LeaciPrixBinary,roy_orbital_2004}, this can be expressed as
\begin{equation}
  \label{eq:24}
  \frac{\Rorb}{c} = \asini\left[ \sin\argp ( \cos E - \ecc) + \cos\argp\sin E \sqrt{1 - \ecc^2} \right],
\end{equation}
where the eccentric anomaly $E(\tau)$ is given by the transcendental equation
\begin{equation}
  \label{eq:25}
  \tau - \tperi = \frac{\porb}{2\pi}\left( E - \ecc\,\sin E\right),
\end{equation}
and the emission time $\tau(t)$ itself depends on $\Rorb$ via Eqs.~\eqref{eq:4},\eqref{eq:19b}.
This set of equations is therefore solved numerically%
\footnote{More details can be found in the documentation of
  \href{https://github.com/lscsoft/lalsuite/blob/master/lalpulsar/lib/SSBtimes.c}{\code{XLALAddBinaryTimes()}}.}.

\section{Mismatch due to timing errors}
\label{sec:mismatchcalculation}

The timing model $\tau(t)$ of Eq.~\eqref{eq:4} relates the detector-frame phase $\phi(t)$ of
Eq.~\eqref{eq:fullSignalModel} to the corresponding source-frame phase $\phi\src(\tau(t))$, and is needed to construct
matched-filter detection statistics such as the $\F$-statistic \citep{jks1}.

Inaccuracies $\dtau(t)$ in the timing model can therefore result in a loss of (recovered) signal power, as these
correspond effectively to a template waveform that is mismatched with respect to the signal.
Namely, from the phase relation Eq.~\eqref{eq:4}, the phase error $\delta\phi$ due to a timing error $\dtau(t)$ can be expressed
as
\begin{equation}
  \label{eq:1}
  \delta\phi(t) = 2\pi\,f\src(\tau)\,\dtau(t) + \Ord{\delta\tau^2}\,,
\end{equation}
where $f\src(\tau)\equiv{d\phi\src(\tau)}/(2\pi{d\tau})$ is the instantaneous source-frame signal frequency, which in
the following we assume to be approximately constant, i.e., $f\src\approx f$.

In order to estimate the corresponding loss in matched-filter signal power, we first construct a simplified
constant-amplitude detection statistic.

\subsection{Simplified detection statistic}
\label{sec:simpl-detect-stat}

The measured data timeseries in a detector is $x_j\equiv x(t_j)$ at sampling timesteps $t_j$ with
$j=1,\ldots N$. For a gravitational signal $h(t)$ embedded in additive stationary Gaussian detector noise $n$
\begin{equation}
  \label{eq:8}
  x_j = n_j + h_j\,.
\end{equation}
The probability density function for $n$ is
\begin{equation}
	\Gauss(n) = k \,e^{-\frac{1}{2}\scalar{n}{n}}\,,
\end{equation}
with zero mean and covariance $\Sigma_{jl}\equiv\expect{n_j n_l}$, with $k$ being a normalisation constant. We use the
scalar-product notation
\begin{equation}
  \scalar{x}{y} \equiv \sum_{jl}^{N} x_j\, \Sigma_{jl}^{-1} y_l\,.
  \label{eq:scalarproduct-a}
\end{equation}
Using the equations above we can write the likelihood for the data under the signal
hypothesis as $\prob{x}{h}=\Gauss(x-h)$, and for the noise hypothesis as $\prob{x}{h=0}=\Gauss(x)$,
resulting in the (log) likelihood ratio
\begin{equation}
  \ln \Lambda(x) \equiv \ln \frac{\prob{x}{h}}{\prob{x}{h=0}} = \scalar{x}{h} - \frac{1}{2}\scalar{h}{h}\,.
  \label{eq:loglikelihood-a}
\end{equation}
Here we only consider narrow-band signals $h$, where only the noise in the narrow
signal band can affect the scalar product Eq.~\eqref{eq:scalarproduct-a} \cite{jks1,finn_detection_1992}.
We further assume the noise to be white over this narrow frequency band, i.e., uncorrelated in the
time domain, namely
\begin{equation}
  \Sigma_{jl} \equiv \expect{n_j\,n_l} \approx \sigma^{2}\delta_{jl},
  \label{eq:10}
\end{equation}
resulting in the simpler scalar-product expression
\begin{equation}
  \scalar{x}{y} = 2\dfact \avg{xy}\,,
\end{equation}
where the data factor $\dfact$ and the standard time-average $\avg{\cdot}$ are defined as
\begin{equation}
  \label{eq:2}
  \dfact\equiv \frac{N}{2\sigma^2}\,,\quad
  \avg{xy} \equiv \frac{1}{N} \sum_{j=1}^N x_j\,y_j\,,
\end{equation}
in analogy to the continuous-time formulation in \citep{prix_analytic_2025}.
The factor of two in the $\gamma$ denominator accounts for the single-sided power-spectral density used in the
continuous-time version.

We further simplify the problem by neglecting the signal amplitude modulation in Eq.~\eqref{eq:fullSignalModel} stemming
from the antenna-pattern functions $F_+$ and the $F_\times$, as timing errors only affect the phase, thus yielding the
constant-amplitude signal model
\begin{equation}
  h(t) = \As\sin \phi(t) + \Ac\cos\phi(t),
  \label{eq:13}
\end{equation}
following the approach of \cite{prix_global_2005}.
For this signal model we obtain
\begin{equation}
  \label{eq:3}
  \scalar{x}{h} = 2\dfact\left( \As \xs + \Ac \xc \right)\,,
\end{equation}
where
$\xs \equiv \avg{x \sin\phi}$, $\xc \equiv \avg{x \cos\phi}$.  We further find
\begin{equation}
  \label{eq:5}
  \scalar{h}{h} = 2\dfact \avg{h^2} \approx \dfact\left(\As^2+\Ac^2\right)\,,
\end{equation}
where averages over many phase cycles are approximated
as $\avg{\sin^2\phi}\approx\avg{\cos^2\phi}\approx 1/2$, and $\avg{\sin\phi\cos\phi}\approx0$.

We can now write the likelihood ratio of Eq.~\eqref{eq:loglikelihood-a} as
\begin{equation}
  \label{eq:6}
  \ln \Lambda(x) = 2\dfact\left(\As\xs + \Ac\xc\right) - \frac{\dfact}{2}\left(\As^2+\Ac^2\right)\,,
\end{equation}
which can be analytically maximized over the two unknown amplitude parameters $\{\As,\Ac\}$, yielding
\begin{equation}
  \label{eq:7}
  \Fphi(x) \equiv \max_{\{\As,\Ac\}}\ln\Lambda = 2\dfact\left(\xs^2+\xc^2\right)\,,
\end{equation}
which is a two degrees-of-freedom version of the classic four degrees-of-freedom $\F$-statistic
\cite{jks1} for the full signal model including amplitude modulation.
This ``constant-response'' $\Fphi$-statistic agrees with Eq.~(9) in \cite{prix_global_2005} and is also discussed in the
appendix of \cite{covas_improved_2022}.

Assuming the data contains a signal $s$, such that $x_{j} = n_{j} + s_{j}$, with zero-mean noise $n$, the expectation
values of the two scalar products are $\expect{x\sub{s,c}} = s\sub{s,c}$, and their variance is
\begin{align}
  \label{eq:9}
  \expect{n\sub{s}^2} &= \frac{1}{N^2}\sum_{jk} \expect{n_jn_k} \sin\phi_j\sin\phi_k\notag\\
                 &= \frac{\sigma^2}{N^2} \sum_j \sin^2\phi_j \approx \frac{\sigma^2}{2 N} = \frac{1}{4\dfact}\,,
\end{align}
using the white-noise assumption Eq.~\eqref{eq:10} and the data factor $\dfact$ defined in Eq.~\eqref{eq:2}.

Both $\sqrt{4\dfact}\,x\sub{s,c}$ are Gaussian distributed with unit variance and means $\sqrt{4\dfact}\,s\sub{s,c}$,
and we see from Eq.~\eqref{eq:7} that $2\Fphi$ is therefore $\chi^2$-distributed with two degrees of freedom and
expectation
\begin{equation}
  \label{eq:11}
  \expect{2\Fphi} = 2 + \rho^2\,,
\end{equation}
with the \emph{signal power} $\rho^2$ in the template with phase $\phi(t)$ defined as
\begin{equation}
  \label{eq:12}
  \rho^2 \equiv 4\dfact(s\sub{s}^2+s\sub{c}^2) = 4\dfact\left|\avg{s\,e^{-i\phi}}\right|^2\,.
\end{equation}

\subsection{Relative loss of signal power}
\label{sec:relative-loss-signal}

Assuming a signal of the form Eq.~\eqref{eq:13} with phase $\phi'(t)$, which we can write as
\begin{align}
  \label{eq:14}
  s(t) &= \As\sin\phi' + \Ac\cos\phi'\notag\\
       &= \frac{A}{2}\left(e^{i(\phi' + \varphi_0)} + e^{-i(\phi'+\varphi_0)}\right)\,,
\end{align}
where $A^2 \equiv \As^2 + \Ac^2$ and $\tan\varphi_0\equiv-\As/\Ac$, then Eq.~\eqref{eq:12} yields the
signal power in the template as
\begin{equation}
  \label{eq:15}
  \rho^2 \approx \rho_0^2 \,\left|\avg{e^{i\delta\phi}}\right|^2\,,
\end{equation}
where
\begin{equation}
  \label{eq:17}
  \rho_0^2 \equiv \scalar{s}{s} = \dfact\left(\As^2+\Ac^2\right)\,,
\end{equation}
is the perfect-match signal power, and
\begin{equation}
  \label{eq:16}
  \delta\phi(t) \equiv \phi'(t) - \phi(t)\,,
\end{equation}
is the phase error between signal and template.
Here we used the fact that for small phase-errors, $\phi'\approx\phi$ and therefore $\avg{e^{i(\phi'+ \phi)}}\approx 0$.

We define the \emph{mismatch} $\mu$ between the signal and the template as the relative loss of signal power
\cite{prix_search_2007}, namely
\begin{equation}
  \label{eq:18}
  \mu \equiv \frac{\rho_0^2 - \rho^2}{\rho_0^2} = 1 - \left|\avg{e^{i\delta\phi}}\right|^2,
\end{equation}
and Taylor-expanding in small $\delta \phi$ yields
\begin{equation}
  \mu = \avg{\delta\phi^{2}} - \avg{\delta\phi}^{2} + \Ord{\delta\phi^4}\,.
\end{equation}
In the more common context of parameter-space metrics \cite{owen_search_1996}, the phase error is due to an offset
$\delta\dop$ in phase-evolution parameters, namely $\delta\phi=(\partial\phi/\partial\dop^i)\,\delta\dop^i +\Ord{\delta\dop^2}$, resulting
in the well-known phase metric \cite{brady_searching_1998,prix_search_2007}, namely
$g_{ij}\equiv \avg{\partial_i\phi\,\partial_j\phi} - \avg{\partial_i\phi}\avg{\partial_j\phi}$, with corresponding
mismatch $\mu = g_{ij}\,\delta\dop^i\delta\dop^j+\Ord{\delta\dop^3}$.

Here we consider instead a phase error $\delta\phi$ that is due to timing errors $\dtau$ via Eq.~\eqref{eq:1}, resulting
in the mismatch expression
\begin{equation}
  \label{eq:mismatch}
  \mu = (2 \pi f)^2\,\stddtau^{2} + \Ord{\dtau^3}\,,
\end{equation}
where we defined the standard deviation $\stddtau$ of the timing errors $\dtau(t)$, i.e.,
\begin{equation}
  \label{eq:23}
  \stddtau^2 \equiv \avg{\dtau^{2}} - \avg{\dtau}^{2}\,.
\end{equation}
To leading order, the mismatch $\mu$ due to timing errors is proportional to their {\it{variance}}, and we can write
\begin{equation}
  \label{eq:mismatch-scaling}
  \mu \simeq \num{4e-5}
  \left(\frac{f}{\SI{1000}{\Hz}}\right)^2
  \left(\frac{\stddtau}{\SI{1}{\micro\second}}\right)^2\,.
\end{equation}

\subsection{Numerical validation of the mismatch prediction}
\label{sec:numer-test-mism}

We validate the robustness of the mismatch prediction Eq.~\eqref{eq:mismatch} by signal injections in data without
noise and using the $\F$-statistic with matched phase-evolution parameters to recover the injected signals.
In order to generate timing errors $\dtau(t)$ between injections and $\F$-statistic templates, we use two (widely)
different ephemeris versions, namely \texttt{DE430} for the injections and \texttt{DE200} for the mismatched
$\F$-statistic, yielding $\rho^2$.
To compute the perfectly-matched $\F$-statistic for $\rho_0^2$ we use \texttt{DE430} again, and the resulting mismatch
$\mu$ is then given by Eq.~\eqref{eq:18}.

We perform these injections and recoveries at \num{20} different signal frequencies, drawn uniformly from
$f\in[40, 3200]\,\si{Hz}$, \num{80} sky-positions uniformly sampled in $\alpha\in[0, 2\pi]$ and
$\delta\in[-\pi/2,\pi/2]$, and for each point we use \num{6} randomly-drawn amplitude-parameter pairs $\{\cosi,\psi\}$
for the injections.
These tests were performed for the L1 detector, with data spanning 237~days from the start of
O4 and an SFT timebase of $T\sub{sft}=\SI{600}{\second}$. 

\begin{figure}[htbp]
  \centering
  \includegraphics[width=\columnwidth]{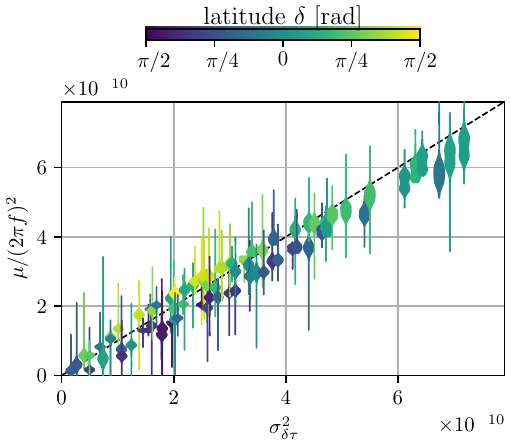}
  \caption{Violin plot showing the distribution of rescaled mismatches $\mu/(2\pi f)^2$ versus timing-error
    variance $\stddtau^2$, generated by injecting signals using \texttt{DE430} ephemeris and recovering them with the
    $\F$-statistic using an offset ephemeris version \texttt{DE200}.
    Each value of $\stddtau^2$ corresponds to one of the \num{80} randomly-drawn sky positions, with the distribution of
    mismatches over random signal frequencies and amplitude parameters $\{\cosi,\psi\}$.
    The dashed black line represents the mismatch prediction of Eq.~\eqref{eq:mismatch}.  }
  \label{fig:mm-injection}
\end{figure}

The resulting mismatches span about six orders of magnitude, $\mu\sim[\num{1e-6}, 0.3]$.
In order to illustrate the predictive power of Eq.~\eqref{eq:mismatch}, in Fig.~\ref{fig:mm-injection} we plot the
distribution (over the 6 amplitudes $\times$ 20 frequencies) of the rescaled mismatch $\mu/(2\pi f)^2$ versus timing-error
variance $\stddtau^2$ (each value corresponding to one sky position).
We see that the measured mismatch is generally in good agreement with the prediction.
We also note a weak dependency of the mismatches on signal latitude $\delta$, namely (for this setup) at similar
$\stddtau^2$ the mismatches at higher latitude tend to be slightly larger than at lower latitudes.
This effect is not captured by our prediction Eq.~\eqref{eq:mismatch}, as the derivation in
Sec.~\ref{sec:simpl-detect-stat} neglects antenna patterns.

\section{Comparison of \lalsuite{} with \pint{}}
\label{sec:comp-lals-with}

In order to validate the \lalsuite{} timing model, we compare it against \pint{} \cite{pint,susobhanan_pint_2024}
(version 1.1.5), a state-of-the-art software package for high-precision pulsar timing.
\pint{} itself was validated against independent timing-model implementations in \tempo{} and \tempoTwo{}, see
\cite{pint,alam_nanograv_2020}.
The numerical timing comparisons reported below were generated with \lalsuite{} commit
\code{84e7563cfc} (\software{LALPulsar} version 3.1.0.1) and \code{pyCW} commit \code{4ba7705ea2}
(version 0.1.5.dev20260813).

We define the disagreement $\dtau$ between the \lalsuite{} time delay $\Delta\super{\lalsuite}$ and \pint{}'s
$\Delta\super{\pint}$ as
\begin{equation}
  \label{eq:disagreement}
  \dtau \equiv \Delta\super{\lalsuite} - \Delta\super{\pint}\,,
\end{equation}
and similar for individual timing-model components $\Delta_C$ (discussed in Sec.~\ref{sec:time-delays}), defining
per-component disagreements $\dtau_C$.

The results shown here use JPL ephemeris version \code{DE405} \cite{jpl} for the Earth and Sun, and we also tested
\code{DE200}, \code{DE421}, \code{DE430}, \code{DE435}, \code{DE436} and \code{DE440}, yielding virtually identical
results.

\subsection{Shapiro delay when crossing the Sun}
\label{sec:shap-delay-cross}

\begin{figure}[htbp]
  \centering
  \includegraphics[width=\columnwidth]{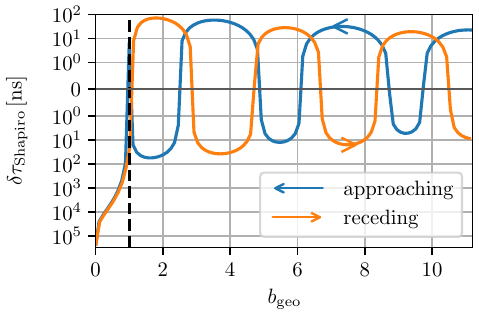}
  \caption{Shapiro-delay disagreement $\dtau\sub{Shapiro}$ versus relative (geocentric) impact parameter
    $b\sub{\geo}$ for a sky position and a timespan of six days around an instant where the signal
    ray passes very close ($\sim\SI{100}{\meter}$) to the center of the Sun.
    The vertical axis is linear within $\pm\SI{1}{\nano\second}$ and logarithmic in magnitude outside this range, with
    the sign retained.
    The left- and right-arrow lines show the approaching and receding time segments, respectively.
    The dashed vertical line at $b\sub{geo}=1$ indicates the surface of the Sun.
    At the center of the Sun \pint's $\Delta\sub{Shapiro}$ diverges.  }
  \label{fig:shapiro}
\end{figure}

We first consider the differences in Shapiro delay for signals crossing the Sun, as discussed in Sec.~\ref{sec:shapiro}.
For this we use a single sky position and a timespan of six days centered on a GPS mid-time $t_0$ for which a signal
arriving at the LIGO Livingston (L1) observatory would pass very close\footnote{Sky position
  $\alpha=\SI[round-mode=places, round-precision=11]{5.792032167516225}{\radian}$,
  $\delta = \SI[round-mode=places, round-precision=11]{-0.20168795317485194}{\radian}$, mid-time
  $t_0=\SI{1234567890}{s}$, avoiding perfect Sun-alignment by $\sim\SI{100}{\meter}$ to prevent \pint{}'s Shapiro delay
  from overflowing.}
% set the sky-position exactly so that at t0 the signal->detector line goes exactly through the center of the Sun
% t0 = 1234567890
% cat = cw.DataCatalog(max_timespan=[t0-1, t0+1], timebase=2, detectors="L1")
% detstate0 = cat.swig_multi_detector_state_series.data[0].data[0]
% sun_det = detstate0.earthState.se - detstate0.earthState.posNow + detstate0.rDetector
% offset from exact sun-center to avoid pint_shapiro->inf
% sun_det += 85.0/lal.C_SI
% vn = -sun_det / np.linalg.norm(sun_det)
% alpha = 2 * np.pi + float(np.atan2( vn[1], vn[0] ))
% delta = float(np.asin( vn[2] ))
to the center of the Sun.
% see figsources/shapiro-sun.ini for that calculation
The Sun-crossing itself takes about $\SI{13}{\hour}$
%% vorb = 2 * np.pi * lal.AU_SI / lal.YRSID_SI
%% > 29784.735487039747
%% 2 * lal.RSUN_SI / vorb / 3600
%% > 12.988420422098423
and we compute Shapiro delay in steps of $\sim\SI{43}{\minute}$.
The disagreement $\dtau\sub{Shapiro}(t)$ is shown in Fig.~\ref{fig:shapiro} as a function of the (geocentric) relative
impact parameter $b\sub{\geo}$ of Eq.~\eqref{eq:19}.
We see that the disagreement is oscillatory and not symmetrical when approaching and departing the center of the Sun,
due to \pint{} computing Shapiro delay (more accurately) with respect to the detector, while \lalsuite{} is using the
geocenter instead.
Outside the Sun the disagreement is bounded by $|\dtau\sub{Shapiro}| < \SI{100}{\ns}$, then rapidly diverges when
approaching the center of the Sun ($b\rightarrow 0$).
As discussed in Sec.~\ref{sec:shapiro}, this is due to the point-mass Shapiro-delay approximation in \pint{} diverging
in this limit, which is not a problem for electromagnetic signals that cannot pass through the Sun.

In the following tests we therefore filter timesteps where a signal would be crossing the Sun, as the \pint{} comparison
reference is not valid in this case.

\subsection{Comparing solar-system delays}
\label{sec:testing-solar-system}

For the following tests we use the location of the L1 observatory and a one-year timespan starting May 24, 2023 (GPS
\num{1368921618} or MJD \num{60088}), which overlaps with the first year of the O4 LIGO-Virgo-KAGRA observing run. 
The results were similar for the H1 detector. 
Time delays are computed and compared in steps of \SI{1}{\hour}, resulting in a total of $\num{8760}$ timesteps per
signal-parameter-space point.

We first look at the disagreement timeseries $\dtau_C(t)$ for different delay components $C$, using three
randomly-picked sky positions.
Given there are different Einstein-delay implementations available in both \lalsuite{} and \pint{} (cf.~
Sec.~\ref{sec:einstein}), we consider three relevant combinations for comparison, labeled as \oldTDB, \newTDB, and
\newTDBbipm.

\subsubsection{\oldTDB: \code{XLALBarycenterEarth} vs \pint{} \TT(\TAI)}
\label{sec:oldTDB}

\begin{figure*}[htbp]
    \centering
    \includegraphics[width=\textwidth]{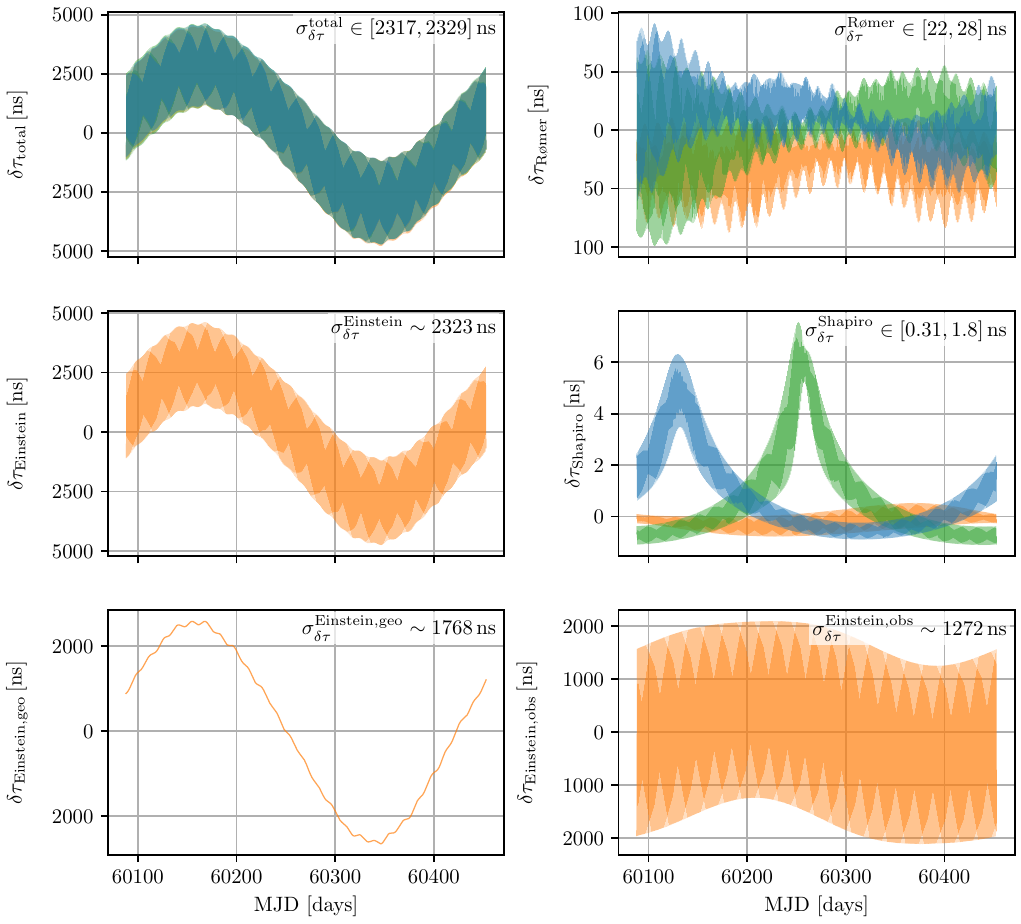}
    \caption{Time-delay disagreements for isolated sources (for three random sky positions, represented by the three colours) over a one-year timespan,
      showing the total delay $\dtau\sub{total}$ (\emph{top left}), \Romer{} delay $\dtau\sub{\Romer}$ (\emph{top
        right}), total Einstein delay $\dtau\sub{Einstein}$ (\emph{middle left}) and Shapiro delay $\dtau\sub{Shapiro}$
      (\emph{middle right}), as well as the geocentric $\dtau\sub{Einstein,geo}$ (\emph{bottom left}) and observatory component
      $\dtau\sub{Einstein,obs}$ (\emph{bottom right}) of the Einstein delay.
      This plot shows the \oldTDB{} comparison between the original \lalsuite{} Einstein-delay implementation
      (\code{XLALBarycenterEarth}) and \pint's \TT(\TAI).
    }
    \label{fig:disagreement_oldTDB}
\end{figure*}
Using the original \lalsuite{} Einstein-delay code path \code{XLALBarycenterEarth}, with \pint{} set to use the same
idealized \TT(\TAI) realization that \lalsuite{} assumes, the resulting disagreement timeseries are shown in
Fig.~\ref{fig:disagreement_oldTDB}.
We see that the total disagreement is dominated by the Einstein delay, with a standard deviation of
$\stddtau\super{Einstein}\sim \SI{2323}{\ns}$, followed by \Romer{} delay $\stddtau\super{\Romer}\sim\SI{25}{\ns}$ and
Shapiro delay $\stddtau\super{Shapiro}\sim\SI{1}{\ns}$.
Note that Einstein delay is independent of sky position, which is why there is only one timeseries for $\dtau\sub{Einstein}$.

Separating the Einstein delay into its geocenter- and observatory components, shown in the last row in
Fig.~\ref{fig:disagreement_oldTDB}, we see that both contribute significantly, namely
$\stddtau\super{Einstein,\geo}\sim\SI{1768}{\ns}$, and $\stddtau\super{Einstein,\obs}\sim\SI{1272}{\ns}$, respectively.
This is consistent with the fact that \code{XLALBarycenterEarth} uses a truncated approximation to the geocentric
Einstein delay and neglects the observatory correction, as discussed in Sec.~\ref{sec:einstein}.

These comparison results agree with the original \tempo{} comparison performed in \cite{LIGO-tempo}, which found that
$|\dtau|< \SI{4}{\micro\second}$, and estimated the resulting mismatch to be of order $10^{-4}$.
Using the observed $\stddtau\sim\SI{2300}{\ns}$ for a signal at $f\sim\SI{1284}{\Hz}$, our prediction
Eq.~\eqref{eq:mismatch} would result in $\mu\sim \num{3.4e-4}$.
%% (2 * np.pi * 1284 * 2300e-9)**2
%% > 0.00034430668598528714
%
This original implementation is therefore (easily) within the design accuracy requirements for the detection of
continuous waves.
For reference, the $\F$-statistic implementation uses various speed-optimizing approximations resulting in mismatches of
order of a few percent.

\subsubsection{\newTDB: \code{XLALBarycenterEarthNew} vs \pint{} \TT(\TAI)}
\label{sec:newtdb}

\begin{figure}[htbp]
    \centering
    \includegraphics[width=\columnwidth]{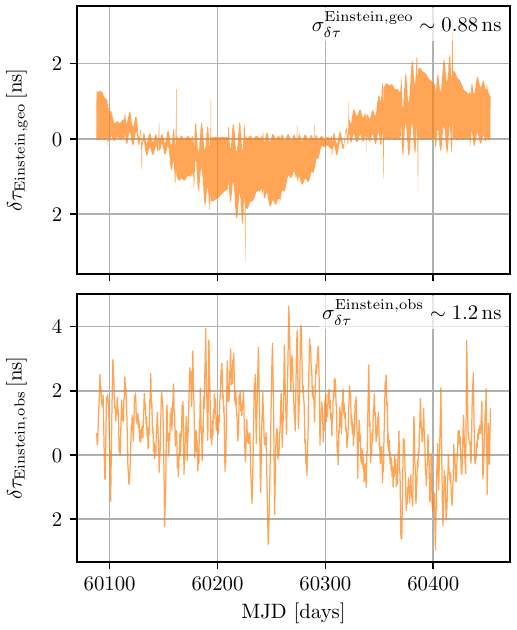}
    \caption{Time-delay disagreements for isolated sources over a one-year timespan, showing the geocentric
      $\dtau\sub{Einstein,geo}$ (\emph{top}) and observatory component $\dtau\sub{Einstein,obs}$ (\emph{bottom}) of the
      Einstein delay.
      This plot shows the \newTDB{} comparison between the newer \lalsuite{} Einstein-delay implementation
      (\code{XLALBarycenterEarthNew}) and \pint's \TT(\TAI).
    }
    \label{fig:newtdb-panel}
\end{figure}

Using the newer Einstein-delay implementation \code{XLALBarycenterEarthNew} in \lalsuite{}, comparing against \pint{}'s
idealized \TT(\TAI) realization, the resulting disagreement timeseries are shown in Fig.~\ref{fig:newtdb-panel}.
We see ns-level agreement in both geocentric- and observatory Einstein delays, resulting in the \Romer-delay
$\stddtau\super{\Romer}\sim\SI{25}{\ns}$ of Fig.~\ref{fig:disagreement_oldTDB} now being the overall bottleneck.

\subsubsection{\newTDBbipm: \code{XLALBarycenterEarthNew} vs \TT(\BIPM)}
\label{sec:newtdbbipm}

\begin{figure*}[htbp]
    \centering
    \includegraphics[width=\textwidth]{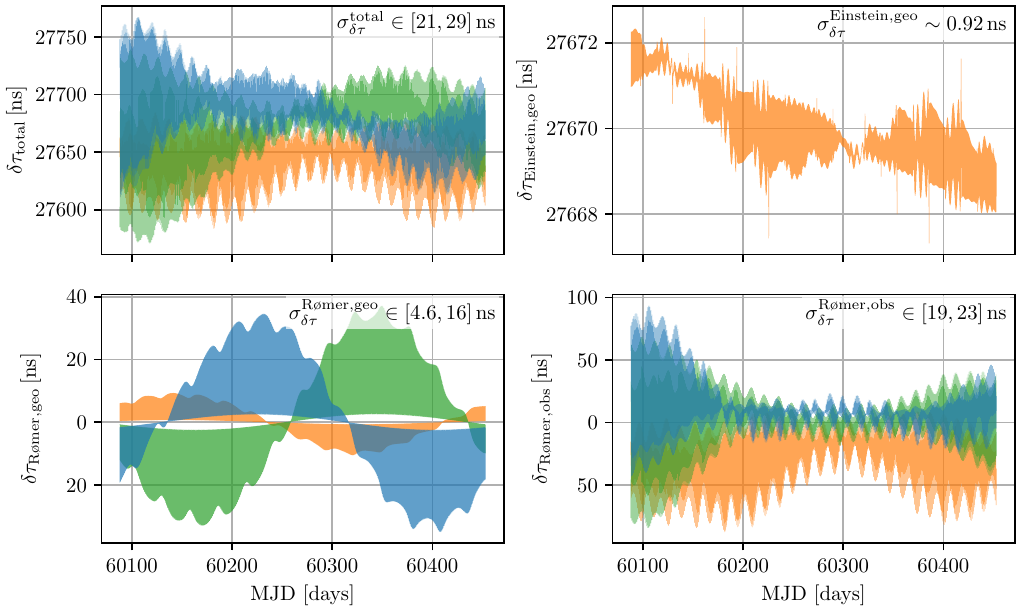}
    \caption{Time-delay disagreements for isolated sources over a one-year timespan, showing the
      total delay $\dtau\sub{total}$ (\emph{top left}), geocentric Einstein delay $\dtau\sub{Einstein,geo}$ (\emph{top
        right}), and the geocentric $\dtau\sub{\Romer,geo}$ (\emph{bottom left}) and observatory $\dtau\sub{\Romer,obs}$
      (\emph{bottom right}) components of the \Romer{} delay.
      This plot shows the \newTDBbipm{} comparison between the newer \lalsuite{} Einstein-delay implementation
      (\code{XLALBarycenterEarthNew}) and \pint's \TT(\BIPM).
    }
    \label{fig:bipm-roemer-panel}
\end{figure*}

Using the more accurate \TT(\BIPM) realization in \pint{}, and comparing against \lalsuite{}
\code{XLALBarycenterEarthNew} (which uses \TT(\TAI)), the resulting disagreement for $\dtau\sub{Einstein,\geo}$ is shown
in the top-right panel of Fig.~\ref{fig:bipm-roemer-panel} (the observatory Einstein delay $\dtau\sub{Einstein,\obs}$ is
not affected by this change).
We see that this results in an overall disagreement shift by about $\SI{27.7}{\micro\second}$, with a slow drift
over time, such that the resulting standard deviation $\stddtau\super{Einstein,\geo}\sim\SI{0.92}{\nano\second}$ (over
one year) is essentially unchanged.
However, depending on the long-term drift between \TT(\TAI) and \TT(\BIPM), this can result in larger variances for
longer observations.

\subsubsection{\Romer{} delay disagreements}
\label{sec:romer-delay-disagr}

When using \code{XLALBarycenterEarthNew}, the total timing-disagreement variation of
$\stddtau\super{total}\sim \Ord{\SI{25}{\ns}}$ is now dominated by the \Romer{} delay, as seen in
Fig.~\ref{fig:bipm-roemer-panel}.
Furthermore, we see that the observatory contribution $\stddtau\super{\Romer,\obs}$ is larger than the geocentric
disagreement $\stddtau\super{\Romer,\geo}$, as confirmed by more extensive Monte-Carlo tests in the next section.
The latter can be traced to the different numerical handling of the JPL ephemerides, while the former stems from the
approximate Earth-rotation model used in \lalsuite{} as opposed to the more accurate empirical IERS Earth rotation model
used in \pint{}, see Sec.~\ref{sec:solar-system-romer}.

\subsubsection{Monte-Carlo sampled comparison}
\label{sec:monte-carlo-sampled}

\begin{figure}[htbp]
    \centering
    \includegraphics[width=\columnwidth]{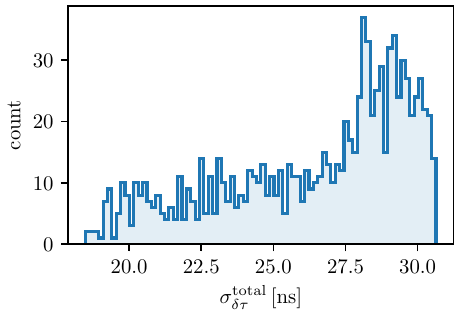}
    \caption{Histogram of standard deviations $\stddtau\super{total}$ of the total timing disagreement over a one-year
      timespan for \num{1000} random sky positions, using the \newTDBbipm{} Einstein-delay comparison.}
    \label{fig:bipm-hist}
\end{figure}

For a more comprehensive comparison, we sample \num{1000} random sky points, for each of the three Einstein-delay test
cases discussed above.
The resulting ranges in standard deviation $\stddtau^C$ for different delay-components $C$ are given in
Table~\ref{tab:summary-solar-system}.
\begin{table}[htbp]
    \centering
    \begin{tabular}{lccc}
\hline
& \multicolumn{3}{c}{[min, max]}\\
$\stddtau~[\si{{\ns}}]$ & \oldTDB & \newTDB & \newTDBbipm \\
\hline
$\dtau\sub{total}$ & [2314, 2332] & [19, 30] & [18, 31] \\
$\dtau\sub{\Romer,\geo}$ & [0.94, 15] & [0.94, 15] & [1.0, 17] \\
$\dtau\sub{\Romer,\obs}$ & [17, 26] & [17, 26] & [17, 26] \\
$\dtau\sub{Shapiro}$ & [0.27, 6.3] & [0.27, 6.3] & [0.27, 6.3] \\
$\dtau\sub{Einstein,\geo}$ & 1768 & 0.88 & 0.92 \\
$\dtau\sub{Einstein,\obs}$ & 1272 & 1.2 & 1.2 \\
\hline
\end{tabular}

    \caption{Measured ranges (over $\num{1000}$ random sky positions) in standard deviations $\stddtau$ of timing
      disagreement, for different delay components (rows) and different comparison variants for the Einstein-delay
      (columns).  }
    \label{tab:summary-solar-system}
\end{table}
As we can see, these ranges are largely consistent with the illustrative results shown above.
In Fig.~\ref{fig:bipm-hist} we show the distribution of $\stddtau\super{total}$ for the most realistic \newTDBbipm{}
comparison case.

We see that overall disagreement variation (using \code{XLALBarycenterEarthNew}) satisfies
$\stddtau\super{total}\lesssim\SI{31}{\ns}$.
Assuming a search at $f=\SI{1000}{\Hz}$, the resulting mismatch Eq.~\eqref{eq:mismatch} would therefore be about
$\mu\lesssim \num{4e-8}$.
%% (2 * np.pi * 1000 * 31e-9)**2
%% > 3.7938759317787484e-08

\subsection{Comparing binary-system delays}
\label{sec:test-binary-syst}

\begin{figure}[htbp]
  \centering
  \includegraphics[width=\columnwidth]{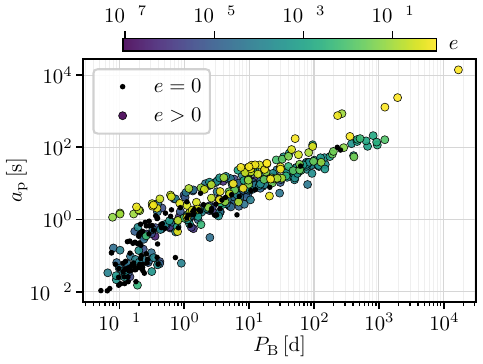}
  \caption{Binary orbital parameters (semi-major axis $\asini$, period $\porb$, and eccentricity $\ecc$) of \num{474}
    binaries from the ATNF catalog, for which we measure the binary-timing disagreements $\dtau\sub{Binary}$.  }
  \label{fig:binary-atnf-parameters}
\end{figure}

In order to test the binary-delay component $\dtau\sub{Binary}$ of Sec.~\ref{sec:binary-delay}, we use the ATNF Pulsar
Catalogue\footnote{Catalogue version 2.8.1, \url{https://www.atnf.csiro.au/research/pulsar/psrcat/index.php?version=2.8.1}}
\cite{manchester_atnf_2005} to generate a realistic test set of binary orbital parameters:
we use the \num{474} binary systems with well-defined periods $\porb$ and semi-major axes $\asini>\SI{e-2}{\second}$
(filtering out a few low-$\asini$ planetary systems producing low-disagreement outliers in the plots), as shown in
Fig.~\ref{fig:binary-atnf-parameters}.

\begin{figure}[htbp]
  \includegraphics[width=\columnwidth]{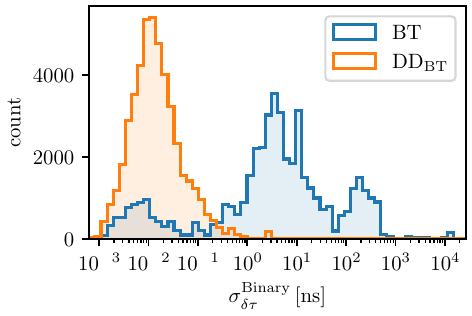}
  \caption{Histograms of standard deviations $\stddtau\super{Binary}$ of the binary timing disagreement
    (between \lalsuite's BT model and \pint's BT and \DD{} models, respectively)
    over a one-year timespan, for the \num{474} binary systems shown in Fig.~\ref{fig:binary-atnf-parameters} with
    \num{100} randomly-sampled values for $\tperi$ and $\argp$ for each system.
  }
  \label{fig:atnf-error-hist}
\end{figure}

\begin{figure}[htbp]
  \centering
  \includegraphics[width=\columnwidth]{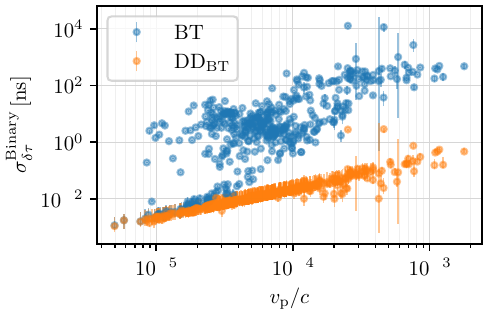}\\
  \includegraphics[width=\columnwidth]{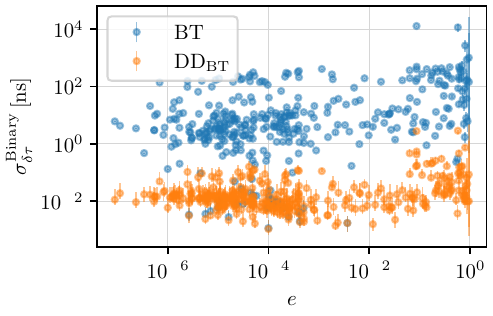}
  \caption{Standard deviations $\stddtau\super{Binary}$ of binary timing disagreement (of \lalsuite's BT versus
    \pint's BT and \DD{} models, respectively)
    versus projected orbital velocity at periapse $v\sub{p}$ (\emph{upper plot})
    and orbital eccentricity $\ecc$ (\emph{lower plot}) for systems with $\ecc>0$.
    Circles mark the median and error bars span the minimum and maximum over the \num{100} random draws of $\tperi$ and
    $\argp$ done per binary system.  }
  \label{fig:binary-atnf-vs-vp-ecc}
\end{figure}

For each of these ATNF binaries, we use their sky position, $\asini$, $\porb$ and eccentricity $\ecc$, and randomly
sample \num{100} realizations for time of periapse $\tperi \in [0, \porb]$ and argument of periapse
$\argp \in [0, 2\pi]$.
The distributions of the resulting disagreements $\stddtau\super{Binary}$ are shown in Fig.~\ref{fig:atnf-error-hist},
comparing either against \pint{}'s BT model or the DD model \cite{ddmodel} with these parameters.
Note, however, that here we are only using the ``BT-subset'' of the DD model, which we denote as \DD.
This comparison therefore does \emph{not} imply that the BT model is an accurate description of all these binaries, only
that the BT-model implementation agrees with \pint{} for these binary parameters.
It is interesting to note that \pint{}'s DD-model implementation seems to agree substantially better with \lalsuite{}
on the BT parameter subset compared to its BT-model implementation.
To further visualize these distributions over the binary parameter space, in Fig.~\ref{fig:binary-atnf-vs-vp-ecc} we
plot $\stddtau\super{Binary}$ versus projected orbital velocity at periapse
$v\sub{p}/c \equiv (2\pi \asini/\porb) \sqrt{(1+\ecc)/(1-\ecc)}$, and versus eccentricity $\ecc$.
This shows a slight upward trend with $v\sub{p}$ and wider distribution spreads when approaching $\ecc\rightarrow 1$,
but otherwise we see consistently very high agreement, typically sub-nanosecond, between \lalsuite{} and \pint's
\DD{} model.

\subsection{Self-consistency check of \lalsuite{} source-time derivatives $\dot{\tau}$}
\label{sec:cons-check-lals}

In addition to the time delays, \lalsuite{} independently computes the source-time derivative
$\dot{\tau}\equiv d\tau/dt=1-\dot{\Delta}$, which is used, for example, in the $\F$-statistic implementation to linearly
interpolate the timing model (computed only once per timestamp) over the timestamp separation (typically the data SFT
timebase of $\lesssim\SI{1800}{\second}$).
We test the internal consistency of the \lalsuite{} source-time derivative $\dot{\tau}$ against a centered
finite-difference estimate obtained from its total time delay $\Delta(t)$, namely
\begin{equation}
  \dot{\tau}\sub{num}(t_i)
  \equiv 1-\frac{\Delta(t_{i+1})-\Delta(t_{i-1})}{t_{i+1}-t_{i-1}}\,.
\end{equation}
We use \num{3000} consecutive timestamps at a spacing of \SI{1}{\second}, starting at GPS time \num{1368921618}, for the
L1 detector and the DE430 ephemeris with the newer TDB Einstein-delay implementation.  The comparison covers \num{200}
isotropically sampled sky positions and orbital phases of binaries with parameters $\asini=\SI{1.4}{\second}$,
$\porb=\SI{68400}{\second}$ and $\ecc=0.1$, with $\tperi$ and $\argp$ sampled randomly.
Over the resulting approximately $6\times10^5$ comparisons, the mean and maximum absolute errors
$|\dot{\tau}\sub{num}-\dot{\tau}|$ are found as
$\SI[per-mode=symbol]{9.5e-12}{\second\per\second}$ and
$\SI[per-mode=symbol]{3.7e-11}{\second\per\second}$, respectively.

\section{Conclusions}
\label{sec:conclusions}

We have performed a systematic validation of the \lalsuite{} timing model for continuous gravitational waves, using
\pint{} as an independent high-accuracy reference implementation, comparing the individual solar-system and binary
time-delay components.
To quantify the impact of timing inaccuracies on searches, we first derived and numerically validated the leading-order
relation Eq.~\eqref{eq:mismatch} between the mismatch $\mu$ and the variance $\stddtau^2$ of the timing errors.

With the original Einstein-delay implementation \code{XLALBarycenterEarth}, total timing disagreement is dominated by
that component and has $\stddtau\simeq\SI{2.3}{\micro\second}$, corresponding to a mismatch of only
$\mu\simeq\SI{0.02}{\percent}$ at a frequency of $f=\SI{1000}{\hertz}$, well within accuracy requirements for
continuous-wave searches.
With the newer Einstein-delay \code{XLALBarycenterEarthNew}, the total disagreement drops to
$\stddtau\lesssim\SI{31}{\nano\second}$, corresponding to $\mu\lesssim\num{4e-8}$ at $\SI{1000}{\hertz}$, see
Sec.~\ref{sec:comp-lals-with} for details.
We tested the \lalsuite{} binary-delay implementation using orbital parameters of \num{474} binary pulsars from the ATNF
catalog and found consistently small disagreements with \pint{}'s \DD{} model of order
$\stddtau\super{Binary}\lesssim\Ord{\si{\nano\second}}$.
We also verified the self-consistency of the independently computed \lalsuite{} source-time derivatives to absolute
errors of $\lesssim\SI[per-mode=symbol]{3.7e-11}{\second\per\second}$.

Finally, we derived \lalsuite{}'s solar-interior Shapiro-delay expression and compared its underlying density model with
a realistic solar model.
While this approximation regularizes the Shapiro delay at the solar center, it underestimates the solar-interior delay
by up to $\sim\SI{14.5}{\micro\second}$ near the center.

These results validate the \lalsuite{} timing model for the tested continuous-wave search
configurations, and provide quantitative information about current accuracy bottlenecks for future improvements, should
they ever be needed.

A standalone python script and \lalsuite{}-patch to reproduce the \pint{}-comparison results can be found at
\url{https://github.com/sharma-kartikey/check-lalsuite-timing-model}.

\begin{acknowledgments}
  We are grateful to Colin J.\ Clark for helping us navigate our way around \pint{} and for useful discussions about
  the intricacies of different Einstein-delay approximations. We thank Heinz-Bernd Eggenstein for pointing out the
  subtleties around accurately predicting the Earth rotation and the IERS Earth rotation models.
 We thank Graham Woan for flagging the limitations of the geometric-optics approximation in our calculation of the
  solar-interior Shapiro delay.

  CC's research was carried out at the Jet Propulsion Laboratory, California Institute of Technology, under a contract
  with the National Aeronautics and Space Administration (80NM0018D0004).

  \textbf{AI tool use}:
  The \href{https://pi.dev}{Pi coding agent} was used with OpenAI's GPT-5.5 and, subsequently, GPT-5.6-sol (both with
  medium reasoning effort) to assist in writing the Python scripts for the \lalsuite{}--\pint{} comparisons and all
  figure-plotting scripts. GPT-5.6-sol also helped identify a realistic solar-density model and implement its numerical
  integration in the Shapiro-delay analysis. AI assistance was further used to polish the Abstract and Introduction and
  to review the manuscript for correctness, identifying several minor errors and inaccuracies. All AI-assisted code and
  text was carefully reviewed, tested, and sanity-checked by the authors, who take full responsibility for the contents
  of the manuscript.
\end{acknowledgments}

\appendix

\section{Solar-interior Shapiro delay derivation}
\label{sec:solar-shapiro-delay}

In this section we document the derivation of the \lalsuite{} Shapiro-delay expressions, namely the standard
solar-exterior solution of Eq.~\eqref{eq:shapiro} and especially the solar-interior expression of
Eq.~\eqref{eq:shapiro-interior}, originally derived by Cutler in 2001 \cite{cutler_behind_sun} and implemented in
\lalsuite{}. We follow the geometric-optics approximation underlying that derivation, which neglects diffraction
and its effects on the observed gravitational-wave phase, as discussed in Sec.~\ref{sec:solar-shapiro-wave-optics}.

\subsection{General Shapiro delay}

Let $U(\vx)$ denote the Newtonian potential, determined by the mass density $\rho(\vx)$ via Poisson's equation,
\begin{equation}
  \nabla^2 U=4\pi G\rho.
  \label{eq:shapiro-newtonian-potential}
\end{equation}
In isotropic coordinates the metric for a static, weak gravitational field (to first order in $U/c^2$) is
\begin{equation}
  ds^2=-\left(1+\frac{2U}{c^2}\right) c^2dt^2
       +\left(1-\frac{2U}{c^2}\right) d\vx^{\,2},
  \label{eq:shapiro-weak-field-metric}
\end{equation}
e.g., Eq.~(4.13) in \cite{schneider_gravitational_1992}.
For a null ray, $ds^2=0$, and expanding to first order therefore gives
\begin{equation}
  dt=\frac{d\ell}{c} - \frac{2U}{c^3}\,d\ell,
  \label{eq:shapiro-null-propagation}
\end{equation}
where $d\ell=|d\vx|$.
The second term denotes the relativistic correction to the flat-space propagation time, and we can therefore write the
general expression for the Shapiro delay as
\begin{equation}
  \Delta\sub{Shapiro}
  =-\frac{2}{c^3}\int_{\mathrm{emission}}^{\mathrm{arrival}}U(\vec{x})\,d\ell,
  \label{eq:shapiro-potential-integral}
\end{equation}
which can be evaluated along the unperturbed straight ray, as corrections from gravitational bending enter at higher
orders of $U/c^2$.

Let $\vro\equiv\vecr-\vecr\sun$ denote the vector from the Sun to the observer.
It is useful to define the observer height $\zo$ and impact vector $\vB$ with respect to the \emph{lens plane}, namely
\begin{equation}
  \zo\equiv\vro\cdot\vn,
  \qquad
  \vB\equiv\vro-\zo\,\vn,
  \label{eq:shapiro-impact-geometry}
\end{equation}
such that $\vB\cdot\vn=0$, and further denote
\begin{equation}
  \ro\equiv\abs{\vro}=\sqrt{\zo^2+B^2},
  \qquad
  B\equiv\abs{\vB}.
  \label{eq:shapiro-impact-magnitudes}
\end{equation}
The unperturbed ray can now be parametrized directly by the axial coordinate $z$, increasing along $\vn$ from
the observer towards the source, namely
\begin{equation}
  \vx(z)=\vB+z\,\vn,
  \quad\text{with}\quad \zo\leq z\leq \zs,
  \label{eq:shapiro-straight-ray}
\end{equation}
where $\vx$ is relative to the solar center, and $\zs$ is the source height over the lens plane.
Using these coordinates we can write Eq.~\eqref{eq:shapiro-potential-integral} as
\begin{equation}
  \Delta\sub{Shapiro}(\vB,\zo,\zs)
  =-\frac{2}{c^3}\int_{\zo}^{\zs}U(\vB,z)\,dz,
  \label{eq:shapiro-thin-lens-integral}
\end{equation}
where $z$ increases from observer to source, therefore $d\ell=|dz|=-dz$ and we reversed the integration direction.

\subsection{Exterior solution}
\label{sec:exterior-solution}

Using the exterior potential for a spherical Sun, i.e., $U(r)=-GM\sun/|\vx|$, with $\abs{\vx(z)}=\sqrt{B^2+z^2}$, we
find
\begin{align}
  \Delta\sub{Shapiro}\super{ext}
  &=\frac{2GM\sun}{c^3}\int_{\zo}^{\zs}\frac{dz}{\sqrt{B^2+z^2}}\notag\\
  &=\frac{2GM\sun}{c^3}
    \ln\left[\frac{\zs+\sqrt{\zs^2+B^2}}
                   {\zo+\sqrt{\zo^2+B^2}}\right].
  \label{eq:shapiro-exterior-finite-source}
\end{align}
Note that the original result by Shapiro \cite{shapiro_fourth_1964} contains an extra term, which is a coordinate
artifact from using Schwarzschild coordinates instead of isotropic ones, see \cite{possel_shapiro_2019} for further
discussion.
For a distant source, $\zs\gg B$, the numerator is approximately constant $\approx2\zs$, corresponding to the unknown
(true) source emission time (see Sec.~\ref{sec:time-delays}).
Following \tempoTwo{}/\pint{} conventions, we set this arbitrary constant to $1\AU$, resulting in
\begin{equation}
  \hspace*{-0.11cm}\Delta\sub{Shapiro}\super{ext}(B,\zo)
  =-\frac{2GM\sun}{c^3}
  \ln\left[\frac{\zo+\sqrt{\zo^2+B^2}}{1\AU}\right],
  \label{eq:shapiro-exterior-impact}
\end{equation}
in agreement with Eq.~\eqref{eq:shapiro}.

\subsection{Interior solution}

For a ray traversing the solar interior, i.e., $\zo<0$ with $B\leq R\sun$ and $|\zo|\sim 1\AU \gg R\sun$, we can
approximate both source and observer as very distant from the lens plane and take
Eq.~\eqref{eq:shapiro-thin-lens-integral} from $\zo\rightarrow-\infty$ to $\zs\rightarrow\infty$.
Applying the transverse two-dimensional Laplacian $\nabla^2_{\!\perp}\equiv\nabla^2 - \partial_z^2$, the integral over
the second term ($\propto\left[\partial_zU\right]_{z=-\infty}^{z=\infty}$) vanishes, and using Poisson's equation
Eq.~\eqref{eq:shapiro-newtonian-potential} we obtain
\begin{equation}
  \nabla_{\!\perp}^2\Delta\sub{Shapiro}(\vB) = -\frac{8\pi G}{c^3}\Sigma(\vB),
  \label{eq:shapiro-thin-lens-pde}
\end{equation}
in terms of the projected surface density, defined as
\begin{equation}
  \Sigma(\vB)\equiv\int_{-\infty}^{\infty}\rho(\vB,z)\,dz.
  \label{eq:shapiro-surface-density}
\end{equation}
This corresponds to the thin-lens formulation Eqs.~(5.13), (4.27a-c) in \cite{schneider_gravitational_1992} (with their
potential $\widehat{\psi}$ related to Shapiro delay as $\Delta\sub{Shapiro}=-\widehat{\psi}/c$).

For an axially-symmetric density distribution, $\Sigma$ and $\Delta\sub{Shapiro}$ only depend on $B$, and
Eq.~\eqref{eq:shapiro-thin-lens-pde} reduces to
\begin{equation}
  \frac{1}{B}\frac{d}{dB}\left(B\frac{d\Delta\sub{Shapiro}}{dB}\right)
  =-\frac{8\pi G}{c^3}\Sigma(B).
  \label{eq:shapiro-thin-lens-ode}
\end{equation}
Integrating this from $0$ to $B$ and assuming regularity at $\Delta\sub{Shapiro}(B=0)$, we obtain
\begin{equation}
  \frac{d\Delta\sub{Shapiro}}{dB} =-\frac{4G}{c^3}\frac{m(B)}{B},
  \label{eq:shapiro-delay-gradient}
\end{equation}
with the cumulative mass $m(B)$ within cylindrical radius $B$ defined as
\begin{equation}
  m(B)\equiv2\pi\int_0^B\Sigma(B')B'\,dB'.
  \label{eq:shapiro-cylindrical-mass}
\end{equation}
A second integration, this time from $B$ to $R\sun$, yields
\begin{equation}
  \Delta\sub{Shapiro}\super{int}(B) = \Delta\sub{Shapiro}\super{ext,\mathit{B=R\sun}}
  + \frac{4G}{c^3}\int_B^{R\sun}\frac{m(B')}{B'}\,dB',
  \label{eq:shapiro-interior-master}
\end{equation}
where we match $\Delta\sub{Shapiro}(B=R\sun)$ to the exterior result of Eq.~\eqref{eq:shapiro-exterior-impact} for a ray
grazing the surface.
This allows us to compute the Shapiro delay for any given axially-symmetric mass distribution $m(B)$ by simple 1D
integration.

\subsection{Solar density-profile models}

In the following we consider three different density profiles, (a) the \lalsuite{} ``singular isothermal-sphere'' model,
(b) a uniform-density model and (c) a realistic tabulated solar density profile.
This comparison serves to illustrate the sensitivity of the Shapiro delay to differences in mass profiles and also
allows us to estimate the error compared to a realistic solar model.

\subsubsection{\lalsuite's singular isothermal-sphere model}

A simple projected surface-density toy model is $\Sigma\propto B^{-1}$, which is inspired by\footnote{This would
  correspond to the projection of a $\rho\propto r^{-2}$ density profile if the integration cutoff is large compared to
  $B$, as assumed for the singular isothermal model in Sec.~8.1.4 of Ref.~\cite{schneider_gravitational_1992}.  } the
singular isothermal-sphere model in \cite{schneider_gravitational_1992}.
Normalizing the mass inside $B\le R\sun$ to be $M\sun$, the \lalsuite{} model is therefore
\begin{equation}
  \Sigma\sub{LS}(B)=\frac{M\sun}{2\pi R\sun B},
  \label{eq:shapiro-sis-surface-density}
\end{equation}
with cumulative mass function $m\sub{LS}(B)=M\sun B/R\sun$.
Defining the relative impact parameter $b\equiv B/R\sun$, Eq.~\eqref{eq:shapiro-interior-master} yields
\begin{equation}
  \Delta\sub{Shapiro}\super{LS}
  = \Delta\sub{Shapiro}\super{ext,\mathit{B=R\sun}}
   +\frac{4GM\sun}{c^3}\left(1-b\right),
  \label{eq:shapiro-sis-interior}
\end{equation}
which is the solar-interior expression of Eq.~\eqref{eq:shapiro-interior} that is implemented in \lalsuite{}.

\subsubsection{Uniform-density sphere}

As an (extreme) example that is less centrally condensed than the \lalsuite{} model, consider a uniform-density sphere,
i.e.,
\begin{equation}
  \rho\sub{Unif}=\frac{3M\sun}{4\pi R\sun^3}.
  \label{eq:shapiro-uniform-density}
\end{equation}
Its projected surface density is
\begin{equation}
  \Sigma\sub{Unif}(B) =\frac{3M\sun}{2\pi R\sun^2}\sqrt{1-b^2},
  \label{eq:shapiro-uniform-surface-density}
\end{equation}
with the corresponding cumulative mass
\begin{equation}
  m\sub{Unif}(b) = M\sun\left(1-(1-b^2)^{3/2}\right),
  \label{eq:shapiro-uniform-cylindrical-mass}
\end{equation}
inserted into Eq.~\eqref{eq:shapiro-interior-master} yields
\begin{equation}
  \begin{aligned}
    \Delta\sub{Shapiro}\super{Unif} &= \Delta\sub{Shapiro}\super{ext,\mathit{B=R\sun}} +\frac{4GM\sun}{c^3} f(b),\quad\text{with}\\
    f(b) &= \frac{4-b^2}{3}\sqrt{1-b^2} - \ln\left(1+\sqrt{1-b^2}\right).
  \end{aligned}
  \label{eq:shapiro-uniform-interior}
\end{equation}
At $b=0$, this interior correction is smaller than the \lalsuite{} model by
\begin{equation}
  \frac{4GM\sun}{c^3}\left(\ln 2-\frac{1}{3}\right)
  \simeq\SI{7.1}{\micro\second}.
  \label{eq:shapiro-uniform-sis-difference}
\end{equation}

\subsubsection{Realistic solar model}

\begin{figure}[t]
  \centering
  \includegraphics[width=\columnwidth]{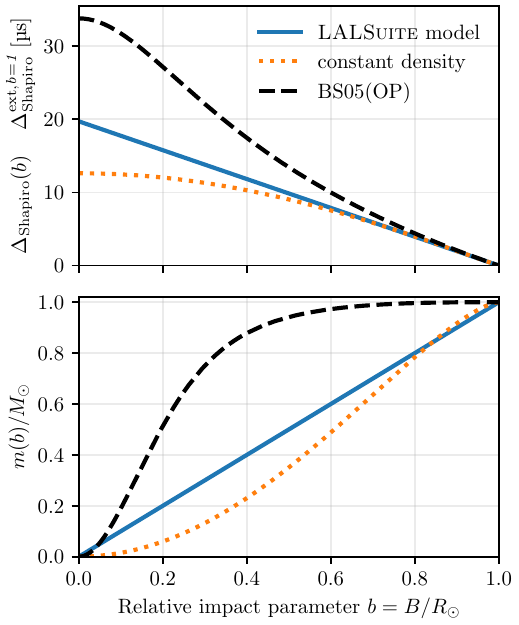}
  \caption{Solar-interior Shapiro delay $\Delta\sub{Shapiro}(b)$ relative to a grazing ray (\emph{upper plot}) and
    cumulative mass function $m(b)/M\sun$ (\emph{lower plot}) versus relative impact parameter $b=B/R\sun$, for three
    different solar-density models.  }
  \label{fig:shapiro-interior-models}
\end{figure}
As a representative realistic reference, we use the tabulated density
profile\footnote{\url{https://www.sns.ias.edu/~jnb/SNdata/Export/BS2005/bs05op.dat}.} for the BS05(OP) solar model
\cite{bahcall_solar_model_2005}.
This density profile is substantially \emph{more} centrally condensed than the \lalsuite{} model except very close to
the center, as seen in Fig.~\ref{fig:shapiro-interior-models} (lower plot), comparing the cumulative mass function
$m(b)$ for the three models considered.
We now numerically integrate the master equation \eqref{eq:shapiro-interior-master} for this density profile, the
resulting interior Shapiro delay $\Delta\sub{Shapiro}\super{BS05(OP)}$ as a function of $b$ is shown in
Fig.~\ref{fig:shapiro-interior-models} (upper plot), comparing all three models.
The larger central mass concentration of the BS05(OP) model yields a larger Shapiro delay compared to the \lalsuite{}
model, by about $\SI{14.5}{\micro\second}$ near the center (at $b\simeq0.045$).

\subsection{Limitations of geometric optics}
\label{sec:solar-shapiro-wave-optics}

As pointed out by G.~Woan \cite{graham}, the above calculations assume geometric optics, while diffraction causes the
observed wave to depend on a finite transverse region rather than a single ray.
For a distant source, the relevant Fresnel radius at the Sun is
\begin{equation}
  r_F = \sqrt{\frac{cD_L}{2\pi f}}
  \simeq 0.38R\sun\left(\frac{f}{\SI{100}{\hertz}}\right)^{-1/2},
\end{equation}
where $D_L\simeq1\AU$ is the distance from the observer to the Sun \cite{takahashi_solar_2023,jung_solar_2023}.
When the ray-based Shapiro delay varies appreciably across a Fresnel-sized region, diffraction modifies the observed
phase, and the corresponding effective delay becomes frequency dependent.
Wave-optics calculations of this effect for continuous waves are found in Sec.~2.3 of \cite{takahashi_solar_2023}.
Woan's estimates \cite{graham} indicate that the geometric-optics comparison between the \lalsuite{} implementation and
the realistic solar model remains approximately valid around $\SI{1}{\kilo\hertz}$, but diffraction becomes important
below a few hundred Hz and reduces the discrepancies to the level of a few microseconds.  

\bibliography{paper.bib}

\end{document}